\documentclass[11pt]{article}
\usepackage{amssymb}
\usepackage[english]{babel}
\usepackage{hyperref}
\usepackage{xcolor}

\newcommand{\be}{\begin{equation}}
\newcommand{\ee}{\end{equation}}
\newcommand{\bea}{\begin{array}}
\newcommand{\ea}{\end{array}}
\newcommand{\beqa}{\begin{eqnarray}}
\newcommand{\eeqa}{\end{eqnarray}}
\newcommand{\bean}{\begin{eqnarray*}}
\newcommand{\eean}{\end{eqnarray*}}

\newcommand{\R}{{\mathbb{R}}}
\newcommand{\eqn}[1]{(\ref{#1})}
\def\up#1{\leavevmode \raise.16ex\hbox{#1}}

\newcommand{\gapproxeq}{\lower
.7ex\hbox{$\;\stackrel{\textstyle >}{\sim}\;$}}
\newcommand{\lapproxeq}{\lower .7ex\hbox{$\;\stackrel
{\textstyle <}{\sim}\;$}}

\newcounter{appendice}

\def\thebibliography#1{{\bf REFERENCES\markboth
{REFERENCES}{REFERENCES}}\list
{[\arabic{enumi}]}{\settowidth\labelwidth{[#1]}\leftmargin\labelwidth
\advance\leftmargin\labelsep
\usecounter{enumi}}
\def\newblock{\hskip .11em plus .33em minus -.07em}
\sloppy
\sfcode`\.=1000\relax}

\begin{document}
\centerline{\LARGE Lagrangian formulation for  electric charge }
\medskip
\centerline{\LARGE in a magnetic monopole distribution}
\vskip .25cm
\bigskip
\bigskip
\centerline{G. Marmo${}^1$\footnote{marmo@na.infn.it},  Emanuela Scardapane${}^1$\footnote{emanuela.scardapane@gmail.com},  A. Stern${}^{2}$\footnote{astern@ua.edu},  Franco Ventriglia${}^{1,3}$\footnote{ventri@na.infn.it} and Patrizia Vitale${}^{1,3}$\footnote{patrizia.vitale@na.infn.it} }

\vskip 0.5cm

\begin{center}
  {${}^1$Dipartimento di Fisica  E. Pancini Universit\`a di Napoli Federico II, \\
Complesso Universitario di Monte S. Angelo, via Cintia, 80126 Naples, Italy
\\}

\end{center}

\begin{center}
  {${}^2$ Department of Physics, University of Alabama,\\ Tuscaloosa,
Alabama 35487, USA\\}

\end{center}

\begin{center}
  {${}^3$INFN Sez. di Napoli \\
Complesso Universitario di Monte S. Angelo, via Cintia, 80126 Naples, Italy
\\}

\end{center}
\vskip 1cm

\normalsize
\centerline{\bf ABSTRACT}
We give  a Lagrangian description of an  electric charge in a field  sourced by a continuous magnetic monopole distribution.  The description is made possible thanks to a doubling of the configuration space.  The Legendre transform of the nonrelativistic Lagrangian agrees with the Hamiltonian description given recently by  Kupriyanov and Szabo\cite{Kupriyanov:2018xji}.  The  covariant relativistic version of the Lagrangian is shown to introduce a new gauge symmetry,  in addition to standard reparametrizations.  The generalization of the system  to 
open strings coupled to a magnetic monopole distribution is also given, as well as  the generalization to  particles in a non-Abelian gauge field which does not satisfy Bianchi identities in some region of the  space-time.

\newpage
\section{Introduction}

It is well known that a local Lagrangian description for an electric charge in the presence of fields  sourced by  an electric charge distribution  requires the introduction of potentials on the configuration space, introducing  unphysical, or gauge, degrees of freedom in the field theory.   If  the field is  sourced  by  a magnetic monopole, the description can be modified by changing the topology of the underlying configuration space, see  e.g.,\cite{Balachandran:1979wx},\cite{Balachandran:2017jha}.
On the other hand, this procedure  has no obvious extension when the fields are sourced by a continuous distribution of magnetic charge.  In that case, auxiliary degrees of freedom can be added, possibly introducing additional local symmetries.   One possibility is to introduce another set of potentials following work of Zwanziger\cite{Zwanziger:1970hk}.
Another  approach is to enlarge the phase space for the electric charge, and this was done recently by Kupriyanov and Szabo \cite{Kupriyanov:2018xji}. The result has implications for certain nongeometric string theories  and their quantization, which leads to nonassociative algebras, see e.g.,\cite{jackiw}-\cite{Szabo:2019hhg}.

The analysis of  \cite{Kupriyanov:2018xji} for  the electric charge in a field  sourced by magnetic monopole distribution is performed in the Hamiltonian setting.   
The  formulation  is made possible thanks to the doubling of the number of phase space variables. 
In this letter we give the corresponding Lagrangian description.  It naturally requires  doubling  the number of configuration space variables.  So here  if $Q$ denotes the original configuration space, one introduces another copy, $\tilde Q$ and writes down dynamics on $Q\times \tilde Q$. While the motion on the two spaces, in general, cannot be separated,  the Lorentz force equations  are recovered when projecting down to $Q$.  The procedure of doubling the configuration space has a wide range of applications, and actually was used long ago in the description of  quantum dissipative systems \cite{bat}-\cite{tHooft:1999xul}.
The description in   \cite{Kupriyanov:2018xji} is
  nonrelativistic.  
 Here, in addition to giving the associated nonrelativistic  Lagrangian, we extend the procedure to the  case of a covariant relativistic particle, as well as to particles coupled to non-Abelian gauge fields that do not necessarily satisfy the Bianchi identity in a region of space-time. As a further generalization we consider the case of an open string coupled to a smooth distribution of magnetic monopoles.  

The outline of this article is as follows.
In section \ref{nonrel} we write down the Lagrangian for  a nonrelativistic   charged particle in the presence of a magnetic field whose divergence field is  continuous and nonvanishing  in a  finite volume of space, and show that the corresponding Hamiltonian description is that of  \cite{Kupriyanov:2018xji}.  The relativistic generalization is given in section \ref{rel}.  Starting with a fully covariant treatment we obtain a new time dependent symmetry, in addition to standard reparametrization invariance.  The new gauge symmetry mixes $\tilde Q$ with $ Q$. Gauge fixing constraints can be imposed on the phase space in order to recover the Poisson structure of the nonrelativistic treatment on the resulting constrained submanifold. Further  extensions of the system are considered in section \ref{further}.  In subsection \ref{further1}  we  write down the action for a particle coupled to a non-Abelian gauge field which does not satisfy  Bianchi identity in some region of   space-time, whereas in    \ref{further2} we generalize to field theory, by considering   an
open string coupled to a magnetic monopole distribution, again violating Bianchi identity.  In both cases we get a doubling of the configuration space variables (which in the case of the  particle in a non-Abelian gauge field includes variables living in an internal space), as well as  a doubling of the number of  gauge symmetries. We note that the doubling of the number of world-sheet degrees of freedom of the string is also the starting point   of  Double Field Theory, introduced by Hull and Zwiebach\cite{HZ},  and further investigated by many authors\cite{DFT}-\cite{dCSV18},  in order to deal with the T-duality invariance of the strings dynamics. This has  its geometric counterpart in Generalized and Double Geometry (see e.g. \cite{GG, gualtieri} and \cite{DG}-\cite{DG5}). Moreover, the doubling of configuration space has also been related to Drinfel'd doubles in the context of Lie groups dynamics \cite{blumen1}-\cite{MPV18} with interesting implications for the mathematical and physical interpretation of the auxiliary variables. 

\section{Nonrelativistic treatment}\label{nonrel}

We begin with a nonrelativistic charged particle on  $ {\mathbb{R}}^3$ in the presence of a continuous magnetic monopole distribution.   Say that the  particle has mass $m$ and charge  $e$  with coordinates and velocities  $(x_i,\dot x_i$) spanning  $T {\mathbb{R}}^3$.  It  interacts with a magnetic field $\vec B(x)$ of nonvanishing divergence $\vec \nabla\cdot \vec B(x)=\rho_M(x).$ In such a case it is possible to show that the dynamics of the particle,  described by the equations of motion
\be\label{eom}
m\ddot x_i=e\epsilon_{ijk} \dot x_j  B_k(x)
\ee
cannot be given by a Lagrangian formulation on the tangent space ${ T {\mathbb{R}}^3}$  because  a vector potential for the magnetic field generated by the smooth monopoles distribution cannot defined,  even locally. (A detailed discussion of this issue will appear in \cite{inprep}.)  On the other hand, a   Lagrangian description is possible if one enlarges the configuration space to $\mathbb{R}^3\times\widetilde {\mathbb{R}}^3$,  and this description leads to Kupriyanov and Szabo's Hamiltonian formulation \cite{Kupriyanov:2018xji}.  For this one extends  the  tangent space to
 $T ({\mathbb{R}}^3\times\widetilde{  {\mathbb{R}}^3})\simeq T {\mathbb{R}}^3\times\widetilde{ T {\mathbb{R}}^3}$.
  We parametrize $\widetilde{ T {\mathbb{R}}^3}$ by   $ (\tilde x_i,\dot{\tilde x}_i), \;i=1,2,3$.  
 A straightforward calculation shows that  the following Lagrangian function
\be L= m{\dot x_i\dot{\tilde x}_i}+e\epsilon_{ijk} B_k(x) \tilde x_i\dot x_j  \;,\label{nrlLgngn}\ee
correctly reproduces Eq. \eqn{eom},  together with  an equation of motion for the auxiliary degrees of freedom $\tilde x_i$
	\be\label{aux}
	 m\ddot{\tilde  x}_i = e\epsilon_{ijk} \dot{\tilde  x}_j  B_k(x) + e\Bigl( \epsilon_{ jk\ell}\frac\partial{\partial x_i}B_k- \epsilon_{ ik\ell}\frac\partial{\partial x_j}B_k\Bigr) \dot x_j\tilde x_\ell
\ee
which are not  decoupled from the motion of the physical degrees of freedom.
Here we do not ascribe any  physical significance to the auxiliary dynamics.  There are analogous degrees of freedom for dissipative systems, and they are associated with  the environment.  Since  our system does not dissipate energy, the same interpretation  does not obviously follow.
The Lagrangian (\ref{nrlLgngn}) can easily be extended to include electric fields.  This, along with the relativistic generalization, is done in the following section.

In passing to the Hamiltonian formalism,  we denote  the momenta conjugate to $x_i$ and $\tilde x_i$ by
\beqa  p_i&=&m\dot {\tilde x}_i-e\epsilon_{ijk} \tilde x_j B_k(x)\cr&&\cr \tilde p_i&=&m\dot x_i
\;, \label{nrcnclmmnt}\eeqa
respectively.  Along with $x_i$ and $\tilde x_i$, they span the 12-dimensional phase space $T^*(\R^3\times \widetilde{\R^3)}$.  The nonvanishing Poisson brackets are \be \{x_i,p_j\}=\{\tilde x_i,\tilde p_j\}=\delta_{ij}\ee

Instead of the canonical momenta (\ref{nrcnclmmnt}) one can define
 \be \pi_i=p_i+e\epsilon_{ijk} \tilde x_j B_k(x)\qquad\quad\tilde \pi_i=\tilde p_i\;,\label{pipitld}\ee
which have the nonvanishing Poisson brackets:
  \beqa \{x_i,\pi_j\}&=&\{\tilde x_i,\tilde \pi_j\}=\delta_{ij}\cr&&\cr
\{\pi_i,\tilde \pi_j\}&=& e \epsilon_{i jk}B_k\cr&&\cr
\{\pi_i,\pi_j\}&=& e\Bigl( \epsilon_{ jk\ell}\frac\partial{\partial x_i}B_k- \epsilon_{ ik\ell}\frac\partial{\partial x_j}B_k\Bigr)\tilde x_\ell\label{kspbs}\eeqa
The Hamiltonian when expressed in these variables  is
\be H=\frac 1m \tilde\pi_i\pi_i\label{ksham}\ee
Eqs. (\ref{kspbs}) and (\ref{ksham}) are in agreement  with the Hamiltonian formulation in  \cite{Kupriyanov:2018xji}. 

 Concerning the issue of the lack of a lower bound for $H$, one can follow the perspective  in \cite{Wilczek}, where a very similar Hamiltonian dynamics is derived.  Namely,  while it is true that  $H$  generates temporal evolution, it cannot be regarded as a classical observable of the particle.  Rather, such  observables should be functions of only the particle's coordinates $x_i$ and its velocities $\tilde \pi_i/m$, whose dynamics is obtained from their Poisson brackets with $H$
\beqa \dot x_i&=&\{x_i,H\}=\frac 1 m \tilde \pi_i \cr&&\cr 
 \dot {\tilde \pi}_i&=&\{\tilde \pi_i,H\}=\frac e m\epsilon_{ijk} \tilde\pi_j B_k  
 \eeqa
The usual  expression for the  energy, $ \frac 1{2m}\tilde \pi_i\tilde \pi_i$,  is, of course, an observable, which is positive-definite and a constant of motion.

\section{Relativistic covariant treatment}\label{rel}

\setcounter{equation}{0}

The extension of the Lagrangian dynamics of the previous section can straightforwardly be made to a   covariant relativistic system. In the usual treatment of a  covariant relativistic particle, written on $T{\mathbb{R}}^4$,  one obtains a first class constraint  in the Hamiltonian formulation which generates  reparametrizations.  Here we find that the relativistic action for a charged particle in a continuous magnetic monopole  distribution, which is now written on $T{\mathbb{R}}^4\times\widetilde{ T{\mathbb{R}}^4}$,  yields an additional first class constraint, generating a new gauge symmetry.  When projecting the Hamiltonian dynamics onto the constrained submanifold of the phase space, and taking the nonrelativistic limit, we recover  the Hamiltonian description of  \cite{Kupriyanov:2018xji}. 

 As stated above,  our action for the charged particle in a continuous magnetic monopole  distribution is written on  $T{\mathbb{R}}^4\times\widetilde{ T{\mathbb{R}}^4}$.  Let us parametrize  $T{\mathbb{R}}^4$ by    space-time coordinates and velocity four-vectors $(x^\mu,\dot x^\mu$), and $\widetilde{ T{\mathbb{R}}^4}$ by   $ (\tilde x^\mu,\dot{\tilde x}^\mu), \;\mu=0,1,2,3$. So here we have included two `time' coordinates, $x^0$ and $\tilde x^0$. Now the dot denotes the derivative with respect to some variable $\tau$ which parametrizes the particle world line in  ${\mathbb{R}}^4\times\widetilde{ {\mathbb{R}}^4}$. The action for a charged particle in an  electromagnetic field $F_{\mu\nu}(x)$, which does {\it not} in general satisfy the Bianchi identity $\frac\partial{\partial x^\mu}F_{\nu\rho}+\frac\partial{\partial x^\nu}F_{\rho\mu}+\frac\partial{\partial x^\rho}F_{\mu\nu}=0$ is
\be
S =\int d\tau \Big\{ m\frac{\dot x_\mu\dot{\tilde x}^\mu}{\sqrt{-\dot x^\nu\dot x_\nu}}+e F_{\mu\nu}(x) \tilde x^\mu\dot x^\nu + L'(x,\dot x) \Big\} \;,\label{flecvrntL}\ee
 $ L'(x,\dot x) $ is an arbitrary function of  $x^\mu$ and   $\dot x^\mu$. Indices are raised and lowered with the Lorentz metric $\eta=$diag$(-1,1,1,1)$. The action is invariant under Lorentz transformations and arbitrary reparametrizations of $\tau$, $\tau\rightarrow \tau'=f(\tau)$, provided we choose $L'$ appropriately. 
 The action is also invariant under a local transformation that mixes  $\tilde {\mathbb{R}}^4$ with  ${\mathbb{R}}^4$,
\be x^\mu\rightarrow  x^\mu \qquad\qquad \tilde x^\mu\rightarrow  \tilde x^\mu+ \frac{\epsilon(\tau)\, \dot x^\mu}{\sqrt{-\dot x_\nu\dot{ x}^\nu}}\;,\label{strngtrnsfm}\ee
for an arbitrary real function $\epsilon(\tau)$.  The first term in the integrand of (\ref{flecvrntL}) changes by a $\tau-$derivative under  (\ref{strngtrnsfm}), while the remaining terms in the integrand are invariant.

 Upon extremizing the action with respect to arbitrary variations $\delta\tilde x^\mu$ of $\tilde x^\mu$, we recover the standard Lorentz force equation on $T{\mathbb{R}}^4$
\be \dot {\tilde p}_\mu=eF_{\mu\nu}(x)\dot x^\nu\;,
\ee
while arbitrary variations $\delta x^\mu$ of $ x^\mu$ lead to
\be \dot{ p}_\mu=e\frac{\partial  F_{\rho\sigma}}{\partial x^\mu}\tilde x^\rho\dot x^\sigma+\frac{\partial L'}{\partial x^\mu}\;\label{thr33ptfr}\ee
  $p_\mu$ and $\tilde p_\mu$ are the momenta canonically conjugate to $x^\mu$ and $\tilde x^\mu$, respectively,
\beqa p_\mu &=&\frac m{(-\dot x^\rho\dot x_\rho)^{3/2}}( \dot x_\mu\dot{\tilde x}_\nu-\dot x_\nu \dot {\tilde x}_\mu)\dot x^\nu-eF_{\mu\nu }\tilde x^\nu+\frac{\partial L'}{\partial\dot x^\mu}\cr&&\cr
\tilde p_\mu&=&\frac{m\dot x_\mu}{\sqrt{-\dot x_\nu\dot{ x}^\nu}}
\label{cnclmmnt}
\eeqa

The momenta $p_\mu$ and $\tilde p_\mu$, along with  coordinates $x^\mu$ and $\tilde x^\mu$,  parametrize a  $16-$dimensional phase space, which we denote simply by  $T^*Q$.   $x^\mu$, $\tilde x^\mu$, $p_\mu$ and $\tilde p_\mu$ satisfy canonical  Poisson brackets relations, the nonvanishing ones being
\be \{x^\mu,p_\nu\}= \{\tilde x^\mu,\tilde p_\nu\}=\delta^\mu_\nu\ee
$\tilde p_\mu$ satisfies the usual mass shell constraint 
\be \Phi_1= \tilde p_\mu\tilde  p^\mu+m^2\approx 0\;,\label{mscnst}\ee
where $\approx$ means `weakly' zero in the sense of Dirac.
Another constraint is 
\be \Phi_2=p_\mu\tilde  p^\mu+e F_{\mu\nu}(x)\tilde  p^\mu \tilde x^\nu\approx 0\;,\label{anthrcnstrnt}\ee  
where from now on we  set $L'=0$.  

{The  three-momenta $\pi_i$ and $\tilde\pi_i$ of the previous section   can easily be generalized to   four-vectors according to
\be 
 \pi_\mu=p_\mu+eF_{\mu\nu}(x) \tilde x^\nu\qquad\qquad\tilde \pi_\mu=\tilde p_\mu\ee
Their nonvanishing Poisson brackets are
\beqa \{x^\mu,\pi_\nu\}&=& \{\tilde x^\mu,\tilde \pi_\nu\}\;=\;\delta^\mu_\nu\cr&&\cr \{\pi_\mu,\tilde \pi_\nu\}&=&e F_{\mu\nu}\qquad  \cr&&\cr \{\pi_\mu, \pi_\nu\}&=&- e\Bigl(\frac\partial{\partial x^\mu}F_{\nu\rho}+\frac\partial{\partial x^\nu}F_{\rho\mu}\Bigr)\tilde x^\rho\label{cvrntpba}\eeqa
Then the constraints 
(\ref{mscnst}) and (\ref{anthrcnstrnt}) take the simple form 
\be \Phi_1= \tilde \pi_\mu\tilde  \pi^\mu+m^2\approx 0\qquad\qquad\Phi_2=\pi_\mu\tilde  \pi^\mu\approx 0\label{cntsntrmspitldpi}\ee  
From (\ref{cvrntpba}), one has $\{ \Phi_1, \Phi_2\}=0$, and therefore
$ \Phi_1$ and $ \Phi_2$
 form a  first class  set of constraints.  They generate the two gauge (i.e., $\tau-$dependent) transformations   on $T^*Q$.   Unlike in the standard  covariant treatment of a relativistic particle, the mass shell constraint $\Phi_1$ does not generate reparametrizations.  $\Phi_1$ instead generates the transformations (\ref{strngtrnsfm}), while a linear combination of $\Phi_1$ and $\Phi_2$ generate reparametrizations.  After imposing  (\ref{mscnst}) and (\ref{anthrcnstrnt})  on $T^*Q$, one ends up with a gauge invariant subspace that is   12-dimensional, which is in agreement with  the dimensionality of the nonrelativistic phase space. 

 Alternatively, one can introduce two additional constraints on $T^*Q$ which fix the two time coordinates $x^0$ and $\tilde x^0$, and thus break the  gauge symmetries.  The set of all four constraints would then form a  second class set,  again yielding a  12-dimensional reduced phase space, which we denote by  $\overline{T^*Q}$.  The dynamics  on the reduced phase space is then determined from Dirac brackets and some Hamiltonian $H$.  We  choose $H$ to be
\be  H=p_0=\pi_0-eF_{0i}(x) \tilde x^i\ee
 $p_0$  differs from $\pi_0 $ in the presence of an electric field.}  The latter can be expressed as a function of the spatial momenta $\pi_i$ and $\tilde  \pi_i$, $i=1,2,3$, after solving the constraints (\ref{cntsntrmspitldpi}). The result is 
\be \pi_0=\frac {\pi_i\tilde\pi_i}{\sqrt{\tilde\pi_j^2+m^2}}\;,\ee 
$\pi_0$ correctly reduces to the non-relativistic Hamiltonian (\ref{ksham}) in the limit $\tilde\pi_j^2<<m^2$.

In addition to recovering the non-relativistic Hamiltonian of the previous section,  the gauge fixing constraints, which we denote by $\Phi_3\approx 0$ and $\Phi_4\approx 0$, can be  chosen such that the Dirac brackets  on $\overline{T^*Q}$ agree with the Poisson brackets (\ref{kspbs}) of the nonrelativistic treatment.   For this take 
\be \Phi_3=x^0-g(\tau)\qquad \qquad\Phi_4=\tilde x^0-h(\tau)\;,\label{gfxngcnst}\ee
where   $g$ and $h$ are  unspecified functions of the proper time.  By definition, the Dirac brackets between two  functions $A$ and $B$  of the phase space coordinates are given by 
\be  \{A,B\}_{\tt {DB}}=\{A,B\}-\sum_{a,b=1}^4\{A,\Phi_a\} M^{-1}_{ab}\{\Phi_b,B\}\;,\label{dfofdb}\ee
where $M^{-1}$  is the inverse of the matrix $M$ with elements $M_{ab}=\{\Phi_a,\Phi_b\},\;a,b=1,...,4$.
From the constraints (\ref{cntsntrmspitldpi}) and (\ref{gfxngcnst}) we get
\be M^{-1}= \frac {1}{2{({\tilde\pi^0})}^2} \pmatrix{0&0&-\pi^0&\tilde\pi^0\cr 0&0&2\tilde\pi^0& 0\cr \pi^0&-2\tilde\pi^0&0&0\cr -\tilde\pi^0&0&0&0}
\ee
Substituting into (\ref{dfofdb}) gives
\beqa  \{A,B\}_{\tt {DB}}&=&\{A,B\}- \frac {1}{2{({\tilde\pi^0})}^2} \,\Biggl(\,\pi^0\Bigl(\{A,x^0\} \{ \tilde \pi_\mu\tilde  \pi^\mu,B\}-\{B,x^0\} \{ \tilde \pi_\mu\tilde  \pi^\mu,A\}\Bigr)\cr&&\cr&&\qquad\qquad\qquad\quad-\tilde\pi^0\,\Bigl(\{A,\tilde x^0\} \{ \tilde \pi_\mu\tilde  \pi^\mu,B\}-\{B,\tilde x^0\} \{ \tilde \pi_\mu\tilde  \pi^\mu,A\}\Bigr)\cr&&\cr&&\qquad\qquad\qquad\;\;-2\tilde\pi^0\,\Bigl(\{A, x^0\} \{ \tilde \pi_\mu \pi^\mu,B\}-\{B, x^0\} \{ \tilde \pi_\mu \pi^\mu,A\}\Bigr)\;\Biggr)\eeqa
It shows that the  Dirac brackets  $\{A,B\}_{\tt {DB}}$ and their corresponding Poisson brackets $\{A,B\}$ are equal if  both  functions $A$ and $B$ are independent of  $\pi^0$ and $\tilde \pi^0$. 
We need to evaluate the Dirac brackets on the constrained subsurface, which we take to be $T {\mathbb{R}}^3\times\widetilde{ T {\mathbb{R}}^3}$,  parametrized  by $x_i,\tilde x_i,\pi_i$ and $\tilde\pi_i$, $i=1,2,3$.  It is then  sufficient to compute their Poisson brackets.  The nonvanishing  Poisson brackets of the coordinates of $T {\mathbb{R}}^3\times\widetilde{ T {\mathbb{R}}^3}$
 are:
\beqa && \{x_i,\pi_j\}=\{\tilde x_i,\tilde \pi_j\}=\delta_{ij}\cr
&&\cr
&& \{\pi_i,\tilde\pi_j\}=e\epsilon_{ijk} B_k\cr&&\cr
&&\{\pi_i,\pi_j\}= e\Bigl( \epsilon_{ jk\ell}\frac\partial{\partial x_i}B_k- \epsilon_{ ik\ell}\frac\partial{\partial x_j}B_k\Bigr)\tilde x_\ell+e\Big(\frac\partial{\partial x_i}E_j-\frac\partial{\partial x_j}E_i\Bigr)h(\tau)\;,\eeqa
where  $F_{ij}=\epsilon_{ijk} B_k$, $F_{0i}=E_i$ and we have imposed the constraint $\Phi_4=0$.  These  Poisson brackets agree with those   
of the nonrelativistic treatment, (\ref{kspbs}), in the absence of the electric field.

\section{Further  extensions}\label{further}

\setcounter{equation}{0}

Here we extend the dynamics of the previous sections to    $1)$ the case  of a particle coupled to a non-Abelian gauge field violating  Bianchi identities   and  $2)$ the case of an
open string coupled to a smooth  distribution of magnetic monopoles.   Of course, another extension would be the combination of both of these two  cases, i.e., where
an
open string interacts with a non-Abelian gauge field that does not satisfy the Bianchi identities in some region of the  space-time.   We shall not consider that here.

\subsection{Particle in a non-Abelian magnetic monopole distribution}\label{further1}

Here we replace the underlying Abelian gauge group of the previous sections, with an  $N$ dimensional non-Abelian Lie  group $G$.  We take it  to be compact and connected with a simple Lie algebra. Given a unitary representation $\Gamma$ of $G$, let  $t_A, \;A=1,2,...N$ span  the corresponding representation $\bar\Gamma$ of the  Lie algebra, satisfying $t_A^\dagger=t_A$,   Tr$\,t_At_B=\delta_{AB}$ and  $[t_A,t_B]=ic_{ABC} t_C$, $c_{ABC}$ being totally antisymmetric structure constants.  In Yang-Mills field theory, the  field strengths now take values in $\bar\Gamma$, $F_{\mu\nu}(x)=f_{\mu\nu}^A(x) t_A$. A particle interacting with a Yang-Mills field carries  degrees of freedom $I(\tau)$ associated with the non-Abelian charge, in addition to  space-time coordinates $x^\mu(\tau)$.  These new degrees of freedom live in the internal space $\bar\Gamma$, $I(\tau)=I^A(\tau) t_A$.   Under gauge transformations, $I(\tau)$ transforms as a vector in the adjoint representation of $G$, just as do  the field strengths $F_{\mu\nu}(x)$, i.e., $I(\tau)\rightarrow  h(\tau)  I(\tau) h(\tau)^\dagger,\; h(\tau) \in \Gamma$.

 The standard equations of motion for  a particle in a non-Abelian gauge field were given long ago by Wong.\cite{Wong:1970fu}   They consist of two sets of coupled equations.
One set  is a straightforward generalization of the Lorentz force law
\be \dot {\tilde p}_\mu={\rm Tr}\Bigl( F_{\mu\nu}(x) I(\tau)\Bigr)\dot x^\nu\;,\label{Wongeq}
\ee
where $ {\tilde p}_\mu$ is again given in (\ref{cnclmmnt}).  The other set consists of first order equations describing the precession of $I(\tau)$ in the internal space $\bar\Gamma$.   Yang-Mills potentials are required in order to write these equations in a gauge-covariant way.

 The  Wong equations were derived from action principles using a number of different approaches. The Yang-Mills potentials again play a vital role in  all of the Lagrangian descriptions.  In the approach of co-adjoint orbits, one takes the configuration space to be $Q={\mathbb{R}}^4\times\Gamma$, and
writes\cite{Balachandran:1977ub},\cite{Balachandran:2017jha}  
\be I(\tau)=g(\tau)Kg(\tau)^\dagger \;,\label{gKgdgr}\ee
where $g(\tau)$ takes values in $\Gamma$, and $K$ is a fixed direction in $\bar\Gamma$.  Under gauge transformations, $g(\tau)$ transforms with the left action of the group,  $g(\tau)\rightarrow h(\tau)g(\tau)$, $h(\tau)\in \Gamma$.  The two sets of Wong equations result from variations of the action with respect to  $g(\tau)$ and $x^\mu(\tau)$.

 Now in the spirit of \cite{Kupriyanov:2018xji}  we imagine that there is a region of space-time where the Bianchi identity does not hold, and so the usual expression for the field  strengths in terms of the  Yang-Mills potentials is not valid.  So we cannot utilize the known actions which yield Wong's equations, as they require existence of the potentials. We can instead  try a  generalization of (\ref{flecvrntL}), which doubles the number of space-time coordinates.  This appears, however, to be insufficient.  In order to have a gauge invariant description for the particle, we claim that it is necessary to double the number of internal variables as well.  Thus we double the entire configuration space, $Q\rightarrow Q\times \tilde Q$. Proceeding along the lines of the coadjoint orbits  approach,  we take $\tilde Q$ to be another copy of ${\mathbb{R}}^4\times\Gamma$. Let us denote all the dynamical variables in this case to be $ x^\mu(\tau)$, $\tilde x^\mu(\tau)$, $g(\tau)$ and $\tilde g(\tau)$, where both $g(\tau)$ and $\tilde g(\tau)$ take values in $\Gamma$  and  gauge transformation with the left action of the group,  $g(\tau)\rightarrow h(\tau)g(\tau)$,  $\tilde g(\tau)\rightarrow h(\tau)\tilde g(\tau)$, $h(\tau)\in \Gamma$.  

 We now propose the following gauge invariant action for the particle
\be S =\int d\tau \biggl\{ {\rm Tr}\, K g(\tau)^\dagger \dot { g}(\tau)-  {\rm Tr}\,I(\tau)\dot{\tilde  g}(\tau)\tilde g(\tau)^\dagger+ m\frac{\dot x_\mu\dot{\tilde x}^\mu}{\sqrt{-\dot x^\nu\dot x_\nu}}+{\rm Tr}\,\Bigl( F_{\mu\nu}(x) I(\tau)\Bigr) \tilde x^\mu\dot x^\nu \biggr\} \;,\label{nAactn}\ee
where $I(\tau)$ is defined in (\ref{gKgdgr}).  To see that the action is gauge invariant we note that the first two terms in the integrand  can be combined to: Tr$\,K g(\tau)^\dagger\tilde g(\tau)\frac d{d\tau}\Bigl( \tilde g(\tau)^\dagger g(\tau)\Bigr),\; \tilde g(\tau)^\dagger g(\tau)$ being gauge invariant.
Variations of $\tilde x^\mu$ in the action  yields  the Wong equation (\ref{Wongeq}).
Variations  of $ x^\mu$ in the action
gives a new set of  equations defining motion on the enlarged configuration space
\beqa  &&\dot{ p}_\mu={\rm Tr}\Bigl(\frac{\partial  F_{\rho\sigma}}{\partial x^\mu}I(\tau)\Bigr)\tilde x^\rho\dot x^\sigma\;, \cr&&\cr &&\qquad{\rm where}\quad\quad p_\mu =\frac m{(-\dot x^\rho\dot x_\rho)^{3/2}}( \dot x_\mu\dot{\tilde x}_\nu-\dot x_\nu \dot {\tilde x}_\mu)\dot x^\nu-{\rm Tr}\Bigl(F_{\mu\nu }I(\tau)\Bigr)\tilde x^\nu
\eeqa
These equations are the non-Abelian analogues of (\ref{thr33ptfr}).
The remaining equations of motion result from  variations of the  $g(\tau)$ and $\tilde g(\tau)$  and describe motion in $\Gamma\times \Gamma$.
Infinitesimal variations of  $g(\tau)$ and $\tilde g(\tau)$ may be performed as follows:   For $\tilde g(\tau)$, it is simpler to consider variations resulting from the right action on the group,  $\delta\tilde g(\tau)=i\tilde g(\tau)\tilde \epsilon(\tau)$, $\tilde\epsilon(\tau)\in\bar\Gamma$.   The action (\ref{nAactn}) is stationary with respect to these variations when
\be\frac d{d\tau} \Bigl(\tilde g(\tau)I(\tau)\tilde g(\tau)^\dagger\Bigr)=0\;,\label{ddtgIGd}\ee
thus stating that $\tilde g(\tau)I(\tau)\tilde g(\tau)^\dagger$ is a constant of the motion.
 For $ g(\tau)$, consider variations resulting from the left action on the group,  $\delta g(\tau)=i  \epsilon(\tau)g(\tau)$, $\epsilon(\tau)\in\bar\Gamma$. These variations lead to the equations of motion
\be \dot I(\tau)=\Bigl[ I(\tau),\dot{\tilde g}(\tau)\tilde g(\tau)^\dagger- F_{\mu\nu}(x) \tilde x^\mu\dot x^\nu \Bigl]\label{dotIt}\ee
The consistency of both (\ref{ddtgIGd}) and (\ref{dotIt}) leads to  the following constraint on the motion 
\be\Bigl[ I(\tau),F_{\mu\nu}(x) \Bigl] \tilde x^\mu\dot x^\nu=0\ee
This condition on $TQ\times T\tilde Q$  is a feature of the non-Abelian gauge theory, and is absent from the Abelian gauge theory.

\subsection{Open string coupled to a magnetic monopole distribution}\label{further2}

Finally we generalize the case of a particle  interacting with  a smooth magnetic monopole distribution, to that of a string interacting  with the same monopole distribution. Just as we doubled the number of particle coordinates in the previous sections, we now double the number of string coordinates.  
We note that a doubling of the world-sheet coordinates of the string, originally limited to the compactified coordinates,  also occurs in the context of Double Field Theory,\cite{DFT} with the original purpose of making the invariance  of the dynamics under T-duality a manifest symmetry of the action. The approach has been further extended to strings propagating in so called non-geometric backgrounds \cite{nongeom1},\cite{hull2},\cite{Plauschinn:2018wbo},\cite{Szabo:2018hhh}, which leads to quasi-Posson brackets, violating the Jacobi identity.  The resolution involves a doubling of the world-sheet coordinates,  similar to what happens in the case under study.

 Whereas   the configuration space for a Nambu-Goto string moving in  $d$ dimensions is  ${\mathbb{R}}^d$, which can have indefinite signature, here we take it to be ${\mathbb{R}}^d\times\widetilde{ {\mathbb{R}}^d}$. Denote the  string coordinates for  ${\mathbb{R}}^d$ and $\widetilde{ {\mathbb{R}}^d}$ by $x^\mu(\sigma)$ and $\tilde x^\mu(\sigma)$, $\mu=0,1,...,d-1$, respectively, where $\sigma=(\sigma^0,\sigma^1)$ parametrizes the string world sheet, ${\cal M}$. $\sigma^0$ is assumed to be a time-like parameter, and $ \sigma^1$ a spatial parameter.  In addition to writing down the induced metric $g$ on $T{\mathbb{R}}^d$, \be g_{\tt a b}=\partial_{\tt a} x^\mu \partial_{\tt b} x_\mu\;, \ee  where $ \partial_{\tt a}=\frac \partial{\partial \sigma^{\tt a}}\;,\;\;{\tt a,b,...}=0,1$\,, we define a non-symmetric matrix $\tilde g$ on $T{\mathbb{R}}^d\times\widetilde{ T{\mathbb{R}}^d}$,
\be \tilde g_{\tt a b}=\partial_{\tt a} x^\mu \partial_{\tt b} \tilde x_\mu\; \ee 
For the free string action we propose to replace the usual Nambu-Goto action by  \be S_0=\frac 1{2\pi\alpha'}\int_{\cal M} d^2\sigma {\sqrt{-\det g}}\;g^{\tt a b}\tilde g_{\tt a b}\;,\label{rlstactn}\ee
where $g^{\tt a b}$ denote matrix elements of $g^{-1}$  and $\alpha'$ is the string  constant. 

The action (\ref{rlstactn}), together with the interacting term given below,  is a natural generalization of the point-particle action Eq. \eqn{flecvrntL} because:  \begin{itemize}
\item Just as with the case of the relativistic point particle action in section \ref{rel}, it is relativistically covariant.
\item  Just as with the case of the relativistic point particle action in section \ref{rel}, there is a new gauge symmetry, in addition to reparametrizations, 
$\sigma^{\tt a}\rightarrow{\sigma'}^{\tt a}=f^{\tt a}(\sigma)$, leading to new first class constraints in the Hamiltonian formalism.  This new gauge symmetry mixes  $\tilde {\mathbb{R}}^d$ with  ${\mathbb{R}}^d$.  Infinitesimal variations are given by
\be  \delta x^\mu =0\qquad\quad \delta\tilde x^\mu=\frac{\epsilon^{\tt a}(\sigma)\, \partial_{\tt a} x^\mu}{\sqrt{-\det g}}\;,\label{trnsfmfs}\ee
where  $\epsilon^{\tt a}(\sigma)$ are arbitrary functions of $\sigma$, which we assume vanish at the string boundaries.  This is the natural generalization of the $\tau-$dependent  symmetry transformation (\ref{strngtrnsfm}) for the relativistic point particle.  Invariance of $S_0$ under variations (\ref{trnsfmfs}) follows from: 
\beqa  
\delta S_0&=&\frac 1{2\pi\alpha'}\int_{\cal M} d^2\sigma {\sqrt{-\det g}}\;g^{\tt a b}\partial_{\tt a}x_\mu\partial_{\tt b}\Bigl(\frac{\epsilon^{\tt c} \partial_{\tt c} x^\mu}{\sqrt{-\det g}}\Bigr)\cr&&\cr
&=&\frac 1{2\pi\alpha'}\int_{\cal M} d^2\sigma g^{\tt a b}\biggl( g_{\tt a c}\partial_{\tt b}\epsilon^{\tt c}+  \partial_{\tt a}x_\mu\partial_{\tt b}\partial_{\tt c}x^\mu\epsilon^{\tt c}-\frac {\partial_{\tt b}\det g}{2\det g}\,g_{\tt a c}\epsilon^{\tt c}\biggr)\cr&&\cr
&=&\frac 1{2\pi\alpha'}\int_{\cal M} d^2\sigma \biggl(\partial_{\tt c}\epsilon^{\tt c}+g^{\tt a b}\Bigl(  \partial_{\tt a}x_\mu\partial_{\tt b}\partial_{\tt c}x^\mu
-\frac 12\partial_{\tt c}g_{\tt a b}\Bigr)\epsilon^{\tt c}\biggr)\cr&&\cr
&=&\frac 1{2\pi\alpha'}\int_{\partial{\cal M}} d\sigma^{\tt a}\epsilon_{\tt a}\;,
\eeqa
which vanishes upon requiring $\epsilon_{\tt a}|_{\partial{\cal M}}=0.$

\item The action (\ref{rlstactn}) leads to the standard string dynamics  when projecting the equations of motion  to $\R^d$.  Excluding for the moment interactions, variations of  the action  $S_0$ with respect to $\tilde x^\mu(\sigma)$ away from the boundary $\partial {\cal M}$  give the   equations of motion
\be \partial_{\tt a} \tilde p^{\tt a}_\mu=0\;,\qquad \tilde p^{\tt a}_\mu=\frac 1{2\pi\alpha'}{\sqrt{-\det g}}\;g^{\tt a b}\partial_{\tt b} x_\mu\ee
These are  the equations of motion for a Nambu string.
 In addition to recovering the usual string equations on $ {\mathbb{R}}^d$, variations of  $S_0$ with respect to $ x^\mu(\sigma)$  lead to another set of the equation of motion on $ {\mathbb{R}}^d\times \tilde {\mathbb{R}}^d$
\be \partial_{\tt a} p^{\tt a}_\mu=0\;,\qquad  p^{\tt a}_\mu=\frac 1{2\pi\alpha'}{\sqrt{-\det g}}\,\Bigl\{(g^{\tt a b}g^{\tt cd}- g^{\tt ad}g^{\tt bc}- g^{\tt ac}g^{\tt bd})\,\tilde g_{\tt cd} \partial_{\tt b}x_\mu+g^{\tt ab}\partial_{\tt b}\tilde x_\mu \Bigr\}\ee

\end{itemize}

Of course,  (\ref{rlstactn}) can be used for both a closed string and an open string.
We now include interactions to the electromagnetic  field.  They  occur at the boundaries of an open string, and are standardly expressed in terms of the electromagnetic potential, which again is not possible in the presence of a continuous magnetic monopole charge distribution.  So here we take instead
\be S_I= e\int_{\partial {\cal M}} d\sigma^{\tt a} F_{\mu\nu}(x)\tilde x^\mu\partial_{\tt a}x^\nu\;,\ee
where $F_{\mu\nu}(x)$, is not required to satisfy the Bianchi identity in a finite volume of   ${\mathbb{R}}^d$. We take $-\infty<\sigma^0<\infty$, $0<\sigma^1<\pi$, with $\sigma^1=0,\pi$ denoting the  spatial boundaries of the  string.  Then the boundary equations of motion resulting from variations of $\tilde x^\mu(\sigma)$ in the total action $S=S_0+S_I$ are
\be   \Bigl(\tilde p^1_\mu+e F_{\mu\nu}(x)\partial_0 x^\nu\Bigr)\Big|_{\sigma^1=0,\pi}=0\;,\ee 
which are the usual conditions in  $ {\mathbb{R}}^d$.
 The boundary equations of motion resulting from variations of $ x^\mu(\sigma)$ in the total action $S=S_0+S_I$ give some new conditions in  $ {\mathbb{R}}^d\times \tilde {\mathbb{R}}^d$
\be   \biggl( p^1_\mu+e \Bigl(\frac\partial{\partial x^\mu} F_{\rho\sigma}+\frac\partial{\partial x^\sigma} F_{\mu \rho}\Bigr)\tilde x^\rho\partial_0 x^\sigma+e F_{\mu\nu}\partial_0 \tilde x^\nu\biggr)\bigg|_{\sigma^1=0,\pi}=0\ee

In the Hamiltonian  formulation of the system  $\pi_\mu=p^0_\mu$ and $\tilde \pi_\mu=\tilde p^0_\mu$ are  canonically conjugate to $x^\mu$ and $\tilde x^\mu$, respectively, having equal time Poisson brackets
\be \Bigl\{x^\mu(\sigma^0,\sigma^1)\,,\,\pi_\nu(\sigma^0,{\sigma'}^1)\Bigr\}=\Bigl\{\tilde x^\mu(\sigma^0,\sigma^1)\,,\,\tilde \pi_\nu(\sigma^0,{\sigma'}^1)\Bigr\}=\delta^\mu_\nu\,\delta(\sigma^1-{\sigma'}^1)\;, \ee
for $0<\sigma^1,{\sigma'}^1<\pi$,
with all other equal time Poisson brackets equal to zero.
The canonical momenta are subject to the four constraints:
\beqa    &&\Phi_1=\tilde \pi_\mu\tilde \pi^{\mu}+\frac 1{(2\pi\alpha')^2}\partial_1 x^\mu\partial_1 x_\mu\approx 0\cr&&\cr&&\Phi_2=\tilde \pi_\mu\partial_1 x^\mu\approx 0\cr&&\cr&&\Phi_3=\pi_\mu\tilde  \pi^{\mu}+\frac 1{(2\pi\alpha')^2}\partial_1 x^\mu\partial_1 \tilde x_\mu\approx 0\cr&&\cr&&\Phi_4=\pi_\mu\partial_1  x^\mu+\tilde \pi_\mu\partial_1 \tilde x^\mu\approx 0\eeqa
It can be verified that they form a first class set.  $\Phi_1$ and $\Phi_2$ generate the local symmetry transformations (\ref{trnsfmfs}), while  linear combinations of the four constraints generate reparametrizations.

\section{Conclusions}
We have considered the problem of the existence of a Lagrangian description for the motion of a charged particle in the presence  of a smooth distribution of magnetic monopoles.  The magnetic field does not admit a potential on the physical configuration space. Auxiliary variables are employed in order to solve the problem, following a procedure commonly used to deal with dissipative dynamics. This is the Lagrangian counterpart of  the Hamiltonian problem, addressed in \cite{Kupriyanov:2018xji}, where the Bianchi identity violating magnetic field entails a quasi-Poisson algebra on the physical phase space which does not satisfy Jacobi identity {unless one doubles the number of  degrees of freedom. 
The problem was further extended to the relativistic case, as well as non-Abelian case. In the last section, we  performed the generalization of the relativistic point-particle action \eqn{flecvrntL} to that of an open  string interacting, once again, with  a  Bianchi identity violating magnetic field. In order to circumvent the problem of the lack of a potential vector, the world-sheet degrees of freedom have been doubled analogous to the case in double field theory.  Many interesting issues can be addressed, such as a possible relationship with double field theory,  or  the quantization problem, which relates Jacobi violation to non-associativity of the quantum algebra. We plan to investigate these aspects in a forthcoming publication. 

\bigskip
\noindent{\bf Acknowledgements}. G.M. is a member of the Gruppo Nazionale di Fisica Matematica(INDAM), Italy. He would like to thank the support provided by the Santander/UC3M Excellence Chair Programme 2019/2020; he also acknowledges financial support from the Spanish Ministry of Economy and Competitiveness, through the Severo Ochoa Programme for Centres of Excellencein RD (SEV-2015/0554).

\end{document}